\documentclass[a4paper]{article}

\usepackage{graphicx}
\usepackage{amsmath}
\usepackage{amssymb}
\usepackage{multirow} 
\usepackage{rotating}

\usepackage{dsfont}
\renewcommand{\Re}{\mathds{R}}

\title{\textsc{A control-theoretical methodology \\ for the scheduling problem}}
\author{Carlo A. Furia $\cdot$ Alberto Leva $\cdot$ Martina Maggio $\cdot$ Paola Spoletini}
\date{}

\begin{document}



\maketitle

\begin{abstract}
This paper presents a novel methodology to develop scheduling algorithms.
The scheduling problem is phrased as a control problem, and control-theoretical
techniques are used to design a scheduling algorithm that meets specific
requirements.
Unlike most approaches to feedback scheduling, where a controller integrates
a ``basic'' scheduling algorithm and dynamically tunes its parameters and hence
its performances, our methodology essentially reduces the design of a
scheduling algorithm to the synthesis of a controller that closes the feedback loop.
This approach allows the re-use of control-theoretical techniques to design
efficient scheduling algorithms; it frames and solves the scheduling problem
in a general setting; and it can naturally tackle certain peculiar requirements
such as robustness and dynamic performance tuning.
A few experiments demonstrate the feasibility of the approach on a real-time
benchmark.
\end{abstract}

\section{Introduction}
\label{sec:introduction}

The word \emph{scheduling} refers to the allocation of resources between
different competing tasks. This generic, abstract definition reflects the
pervasiveness of the scheduling concern across disciplinary fields.
A concrete class of scheduling problems is obtained by specifying a type of
system and tasks, and the goals of the scheduling action.

In this paper, we outline a general methodology to tackle the scheduling
problem. Our approach exploits \emph{control theory} to formulate the
scheduling problem and to solve it. The control-theoretical paradigm represents
the interaction between two distinct parts of a system: the \emph{plant} and
the \emph{controller}. The plant represents the part of the system whose 
dynamics is not modifiable directly, and that must be put under control.
The controller, on the other hand, is a component that provides suitable input
to the plant with the goal of influencing its dynamics towards meeting some
requirements. The controller chooses its action according to the output of the
plant, hence the denomination \emph{feedback control}.

The idea of using control theory to solve scheduling problems is not new.
Indeed, the research area of \emph{feedback scheduling} is based on these
premises \cite{LuEtAl-2002a,HellersteinEtAl-2005a}.
The novelty of our approach consists in how the control-theoretical paradigm
is applied to the scheduling problem, and more precisely which parts of the
system are modeled as the plant and as the controller, respectively.

The most common approach to feedback scheduling supplements an existing
scheduler with a control-theoretical model: the plant is the ``basic''
scheduler itself, and the controller tunes its dynamics over time according
to the evolution of the rest of the system. We suggest a different partitioning,
where the controller \emph{is} the scheduler and the plant is a very abstract
model of the pool of tasks and, in some sense, the resources they run on.

Our stance has a couple of significant advantages over the traditional
approaches. First, it allows the effective re-use of an extensive amount of 
powerful results from classical control theory to smoothly design scheduling 
algorithms. Second, it is remarkably flexible and can easily accommodate some 
complex and peculiar scheduling requirements, such as robustness towards 
disturbances, dynamic adjustment of performance, and a quantitative notion of 
convergence rates.

The approach is general and applies to a large class of scheduling problems.
It is naturally applicable to scheduling the CPU in \emph{real-time 
systems} \cite{ButtazzoEtAl-2004a,LiuLayland-1973a}, which are characterized 
by a quantitative treating of time.  As it is common with feedback scheduling,
it focuses on \emph{soft} real-time, where the failure to respect a deadline
does not result in a global failure of the system, and average performance is
what matters.

The heterogeneous scope of the scheduling problem and the sought generality of the present approach make, at times, the presentation of the technical details necessarily abstract: it is impossible to formalize each and every (domain-specific) aspect of the scheduling problem (e.g., deadlines, priorities, granularities, etc.) in a unique model that is practically useful.
Additionally, different formalizations are often possible, and choosing the best one depends largely on application-specific details, such as whether one is dealing with a batch or a hard real-time system.
The overall goal of the present paper is high-level: outlining the framework proposed, formalizing its basic traits, and demonstrating its flexibility with a few examples.
Focusing the framework on specialized classes of scheduling problems and comparatively assessing its performance belongs to future work.

The rest of the paper presents our approach to feedback scheduling and is
organized as follows. Section \ref{sec:motivations} presents some additional
motivation, with a more direct comparison to the literature which is most 
closely related to this paper. Section \ref{sec:methodology} introduces our 
methodology for the scheduling problem; it focuses on presenting the conceptual 
contribution in a general setting. Section \ref{sec:experimental} discusses an 
experimental validation of the approach, where the general methodology is 
instantiated to solve a few specific concrete problems in a real-time 
scheduling benchmark. Finally, Section \ref{sec:conclusions} draws some 
conclusions and outlines future work.

\section{Motivation and related work}
\label{sec:motivations}

Hellerstein et al.'s book \cite{HellersteinEtAl-2004a} is a comprehensive
review of the applications of control theory to computing-system
problems such as bandwidth allocation and unpredictable data traffic management.
In general, control theory is applied to make computing systems adaptive, more 
robust, and stable. Adaptability, in particular, characterizes the response 
required in applications whose operating conditions change rapidly and
unpredictably.

Lu et al.~\cite{LuEtAl-2001a,LuEtAl-2006a} present contributions in this 
context, for the regulation of the service levels of a web server. 
The variable to be controlled is the delay between the arrival time of a request
and the time it starts being processed. The goal is to keep this delay to within
some desired range; the range depends on the class of each request.
An interesting point of this works is the distinction between
the transient and steady state performances, in the presence of variable
traffic. This feature motivates a feedback-control approach to
many computing-system performance problems.

Scheduling is certainly one of these problems where the transient to steady-state 
distinction features strongly. Indeed, many \emph{ad hoc} scheduling approaches
solve essentially the same problem in different operating conditions.
This is one of the main reasons why \emph{feedback scheduling} has received 
much attention in recent years (see Xia and Sun \cite{XiaSun-2006a} for a 
concise review of the topic). As we argued already in the introduction, the 
standard approach in feedback scheduling consists in ``closing some control 
loop around an existing scheduler'' to adjust its parameters to the varying 
load conditions. This may yield performance improvements, but it falls short 
of fully exploiting the rich toolset of control theory.

For example, in Abeni et al.~\cite{AbeniEtAl-2002a}, the controller
adjusts the reservation time (i.e., the time the scheduler assigns to
each task) with the purpose of keeping the system utilization below a
specified upper bound. The plant is instead a switching
system with two different states, according to whether the system can satisfy 
the total amount of CPU requests or not.
Some tests with a real-time Linux kernel show that the adaptation
mechanism proposed is useful for to improve quality-of-service measurements. Continuing
in the same line of work, Palopoli and Abeni~\cite{PalopoliAbeni-2009a} combine
a reservation-based scheduler and a feedback-based adaptation mechanism to
identify the best parameter set for a given workload.
Block et al.~pursue a similar approach \cite{BlockEtAl-2008} where they integrate feedback models with optimization techniques.

In Lawrence et al.~\cite{LawrenceEtAl-2001a}, the controller adjusts the 
reservation time to within an upper bound given by the most frequently 
activated task. The model of the plant is a continuous-time system whose 
variables record the queuing time of tasks.
The effectiveness of the method proposed is validated through simulations.

Lu et al.\ in~\cite{LuEtAl-2002a} consider some basic scheduling policies 
(both open-loop and closed-loop) and design a controller that prevents system 
overloading. Such goal is achieved by letting some tasks out of the queue when 
the system workload is too high.

All these approaches target the same problem: assigning CPU time to a pool of 
tasks to meet some goals. The devised algorithms are usually extremely 
efficient, but their scope of applicability is often limited to a fairly 
specific domain (e.g., periodic processes with harmonic frequencies deadlines).
Moreover, in all the cited approaches the controller modifies the behavior of 
a ``basic'' scheduling algorithm; indeed, the model of the scheduler is often 
combined with (some aspects) of the processor model, even if their functions 
are in principle clearly distinct.
We believe that this lack of separation of concerns is the result of the 
close adherence to a specific scheduling problem domain, and we claim that 
enforcing a stricter separation in the model can result in some distinctive 
advantage.

The rest of the paper presents an approach where the scheduler is solely
responsible for selecting which tasks have to run and their desired execution time.
The scheduler is then built as the controller that meets some requirements for
such a selection. Notice that the homogeneous nature of the controller (i.e., 
the scheduler) and the plant (i.e., the tasks' execution model) is peculiar to 
computer systems, and makes a unitary design of the overall system easier.
The approach itself can be, we believe, more general and flexible than the 
aforementioned others.

\section{The methodology}
\label{sec:methodology}
This section outlines a methodology to tackle the scheduling problem.
For clarity, it is phrased in terms of allocating CPU time to a set of tasks in
a mono-processor operating system. It should be clear, however, that the 
solution refers to a more abstract class of problems and is relatively general.
In the rest of the paper, we assume familiarity with the basic 
control-theoretical terminology and notation (see e.g., 
\cite{HellersteinEtAl-2004a}).

\begin{figure}[t]
  \centering
  \includegraphics[width=\columnwidth]{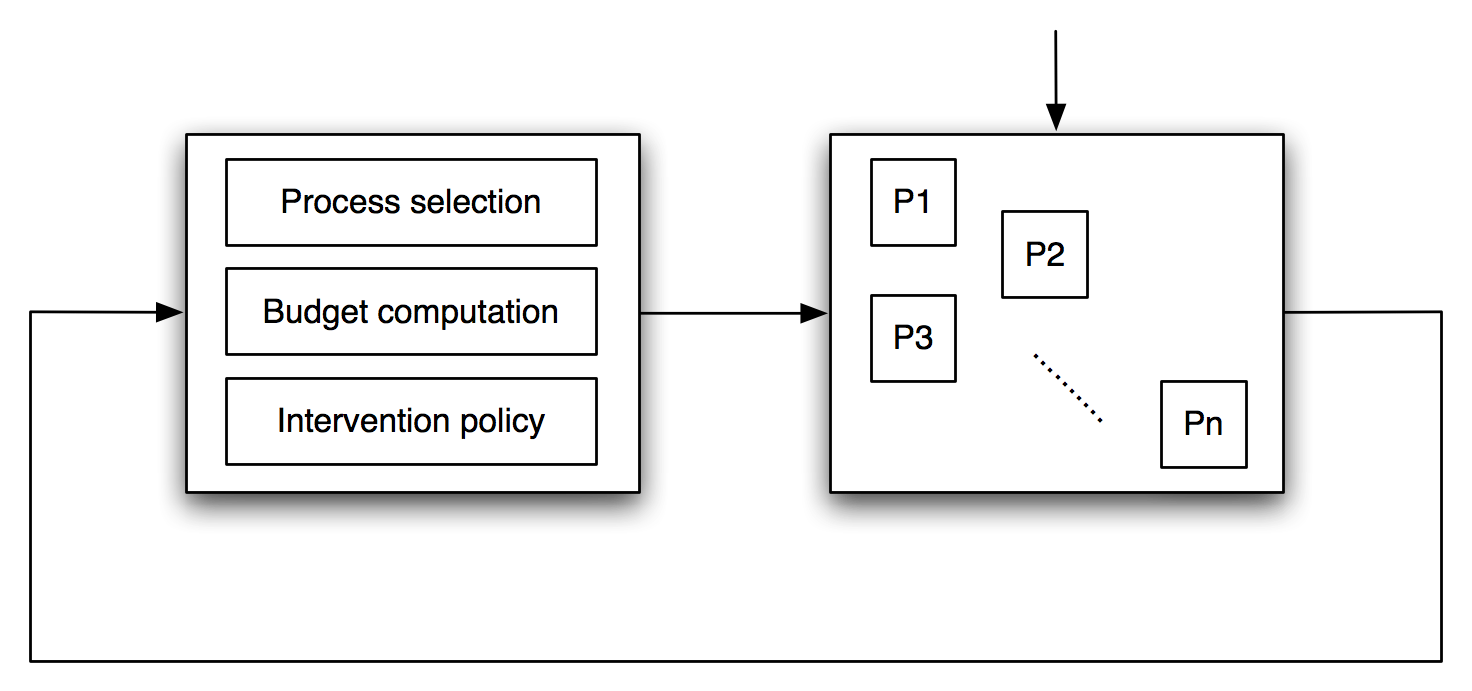}
  \caption{The general framework proposed.}
  \label{fig:framework}
\end{figure}

The basic modeling assumption completely separates the processor and the 
scheduler: the scheduler chooses the order in which the tasks are executed and
their execution times, while the processor actually runs them.
This separation let us focus more precisely on the characteristics of each 
component and understand how to change each model according to the requirements 
we have to meet.
Figure \ref{fig:framework} shows the ``big picture'' of how the scheduling 
problem is cast as a control problem.

The processor is a system that executes tasks according to the input 
received from the scheduler. The scheduler, which provides this input, is then 
the controller of the processor, which is the plant. The control action is 
divided in three aspects: task selection, budget computation, and choice of 
the intervention policy. The first two phases choose which tasks will be 
executed, in what order, and the budget --- defined as the maximum running time before 
preemption --- assigned to each of them.

The intervention policy, instead, determines when the scheduler will next be 
activated. In the following, we assume a straightforward intervention policy
where the scheduler runs after every scheduling round. More complex policies 
can of course be introduced according to the requirements of specific
applications; for example, the scheduler might run whenever the difference
between the desired execution time and the real measured time exceeds a certain
threshold. A detailed analysis of this aspect is orthogonal to the rest of our
methodology and belongs to future work.

Separating the scheduler action in three components facilitates modifications
of the controller model according to different requirements. More precisely,
the overall  model structure remains the same and only the equations modeling 
the affected aspects need to be changed.

Notice that anything that influences the behavior of the running tasks, other
than the scheduler action, is modeled as an exogenous disturbance: an action 
that prevents the system from reaching the goal requirements, which the 
scheduler wants to contrast. This modeling assumption is suitable for factors 
that are, for all practical purposes, unpredictable and unmodifiable.
The notion of disturbance (basically disturbance rejection) from control 
theory is then adopted to model these factors, with the immediate benefit of 
having at our disposal a powerful set of theoretical tools to tackle 
effectively the ensuing problems.

The abstractness and genericity of our framework come with the potential 
drawback of making it difficult to implement the scheduling policies within an 
existing scheduler architecture, which can differ significantly from the 
abstract modular structure of Figure \ref{fig:framework}.
Anyway, we believe that the theoretical analysis that can be carried out
within our framework is extremely useful to determine the criticalities of 
the system under design, even in the cases in which the final implementation 
will require \emph{ad hoc} adjustments.

\subsection{The plant}
\label{sec:plant}
The ``open loop'' model of the plant describes the process executor as a 
discrete-time system. It receives a schedule (which will be the output of the
scheduler described in the next subsection) as input and returns the outcome
of executing the tasks as required by the schedule.

A \emph{round} is the time between two consecutive runs of the scheduler.
Assume that more than one task can be scheduled for execution in a given 
round; correspondingly, we introduce the following variables to describe the 
plant:
\begin{itemize}
\item $N$, the number of tasks to be scheduled;
\item $\tau_p(k)\in \Re^N$, the actual running times of the tasks in the $k$-th
      scheduling round;
\item $\tau_r(k)\in \Re$, the duration of the $k$-th round;
\item $s(k)\in \Re^N$, the schedule at the $k$-th round: an ordered list of the budgets, one for each task; the order determines the execution order and a budget of $0$ means that the task is not scheduled for execution in that round;
\item $\delta b(k)\in \Re^N$, the disturbance during the $k$-th round, 
      defined as the difference between the assigned budget and the actual 
      running time of a task;\\
      (Notice that this variable models uniformly a variety of possible 
      specific phenomena, such as a task that yields control or terminates 
      before preemption, an interrupt occurring, the execution of a critical 
      section where preemption was disabled, etc.)
\item $t \in \Re$, the total time actually elapsed from the system 
      initialization.
\end{itemize}

The model of the plant is then the following system of equations:
\begin{equation}
 \left\{
   \begin{array}{rcl}
    \tau_p(k) &=& s(k-1) + \delta b(k-1)   \\
    \tau_r(k) &=& r_1 \tau_p(k-1)          \\
    t(k)      &=& t(k-1) + \tau_r(k-1) \\
   \end{array}
  \right.
 \label{eqn:BasePlantModel}
\end{equation}
where $r_1$ is a row vector of length $N$ with all unit elements.

Model \eqref{eqn:BasePlantModel} is linear and time-invariant. Negative budgets
are not allowed and, correspondingly, each $s(k)+\delta b(k)$ element cannot be
negative. However, this is irrelevant for the controller, since the set of
considered variables is smaller than the domain limitations. Notice that the
discrete-time model assumes that the scheduler is active only once per round.
Clearly, some $s(k)$ elements can be zero, meaning that not all the tasks will
actually run. The $\tau_r$ variable models round duration, which takes into 
account system responsiveness issues.

\subsection{The scheduler}
\label{sec:scheduler}

A scheduler should usually achieve the following goals, regardless of the 
specificities of the system where it runs \cite{Tanenbaum-2007a}.
\begin{itemize}
\item \emph{Fairness}: comparable tasks should get comparable service;
      (This obviously does not apply to tasks with different properties.)
\item \emph{Policy enforcement}: the scheduler has to comply to general 
      system policies; (This aspect is especially relevant for real-time 
      systems where constraining system policies are usually in place.)
\item \emph{Balance}: all the components of the system need to be used as 
      uniformly as possible.
\end{itemize}

In addition to these general requirements, a scheduler must also achieve 
additional goals that are specific to the system at hand.
For instance, in batch systems, where responsiveness is not an issue, the 
scheduler should guarantee maximization of the throughput, minimization of 
the turnaround time, and maximization of CPU utilization.
In interactive systems, on the contrary, minimization of response time and 
proportionality guarantees are likely scheduling goals.
Finally, deadlines and predictability are specific to real-time systems.

In the following, we outline a general approach to design a scheduler --- based 
on control-theory and the framework presented above --- that achieves a defined
set of goals. Unlike the standard approach that designs a new algorithm for a 
new class of systems, we can accommodate most scenarios within the same
framework by changing details of the equations describing the control model.

\textbf{Process selection and budget computation.}
The scheduler decides which tasks to activate and chooses
a budget for them. This is achieved by setting variable $s_i(k)$ which defines
the budget assigned to the $i$-th task at round $k$.
This action is actually made of two conceptually different parts. A Process
Selection Component (PSC) takes care of deciding the next task to be executed
by the processor, while a Budget Computation Component (BCC) fixes the
duration of the execution for each selected task. If more than one
task is to be executed per round, PSC computes an ordered list of tasks
and BCC assigns one or more budgets to the elements of the list. 
Execution need not be continuous: if a time $\hat{t_i}$ is assigned to the 
$i$-th task, the actual execution can be split into multiple slots within the 
same round.

The distinction between PSC and BCC is modeled by defining $s(k)$ as
$S_\sigma(k)\,b(k)$, where $S_\sigma(k)$ is a $N\times n(k)$ matrix representing
the tasks selected at the $k$-th round, while $b(k)\in \Re^{n(k)}$ represents
the budget assigned to the selected tasks. Notice that, in the most general
case, the number of tasks that can be executed at each round is a 
variable $n(k)$.

PSC can operate statically or dynamically.
In the first case, the strategy is independent of the previous choices,
such as in Round Robin (RR) scheduling. 
In the second case, PSC retains a history of the previous choices and bases its 
new choice on it, such as in fair-share scheduling \cite{KayLauder-1988a}.

In case of static PSC, the matrix $S_\sigma(k)$ is not explicitly function of
$b(i)$ and $S_\sigma(j)$, with $i,j<k$. This means that $S_\sigma(k)$ may not be
a function of both $\tau_p$ and $\tau_r$ in the previous rounds; indeed, these
represent the actual behavior of the CPU with respect to each task. 
Therefore, they may not reflect the choice that the scheduler made
in previous rounds, due to contingencies in the execution of the system.
Consider, as a more concrete example, the shortest remaining time next
algorithm: the PSC chooses the next task to be executed
according to their remaining running times, which obviously depend on what
actually happened in the previous rounds (i.e., the history of $\tau_p$),
but not necessarily on the scheduler's choice (i.e., $s$).

Once the PSC has selected the tasks to be executed, the BCC computes the
budgets for them, by setting $b(k)$. PSC can be static or dynamic, too: in the 
first case the budget is a constant vector $b(k)=\hat{b}$, whereas in the second 
case the budget may change at every round. 

\textbf{Designing the controller.}
Let us now discuss how to define and enforce some of the previously outlined
features in a scheduler, for given PSC and BCC.

\subsubsection{Fairness}
A fair scheduling algorithm has the property that, in every fixed time interval,
the CPU usage is proportionally distributed among tasks, accordingly to their
weights. For the sake of simplicity, let us focus on a fixed number of rounds
$H$.\footnote{Generalizing this approach to deal with a time window, rather 
than a number of rounds, is straightforward.}
Let $p_i(k)$ be the weight 
of the $i$-th task at the $k$-th round. In order to guarantee fairness for
each task, the scheduler must achieve the following equation:
\begin{equation}
 \sum_{i=k}^{k+H}\tau_{p_i(k)} = \sum_{i=k}^{k+H}
                                 \cfrac{p_i(k)}{\sum_{j=1}^{N}p_j(k)} \tau_r(k)
 \label{eqn:eqfair}
\end{equation}
Informally, \eqref{eqn:eqfair} means that the scheduler distributes the CPU
among the different tasks proportionally to their weights (over the next 
$H$ rounds). Then, the scheduler computes a $s(k)$ which satisfies equation 
\eqref{eqn:eqfair}. The algorithm to compute $s(k)$ comes from the solution to the corresponding control problem, for example by means of optimal control theory\footnote{See \cite{Doyle-1996a} for an overview of optimal control theory and further references on the subject.}: find the optimal value of the controlled variable $s(k)$, given a certain cost function $J$ of the state variables $\tau_p, \tau_r, t$.

\subsubsection{Policy enforcement}
The details of how to handle this aspect within our control framework depend
essentially on \emph{which} system policy should be enforced: the term 
``policy'' can refer to very disparate concerns. The experiments described in 
Section \ref{sec:experimental} will tackle a specific instance of activation 
policy.

Let us notice, in passing, that the strict coupling between the system policy 
and the features of the controller that enforce such a policy is one of the 
reasons why most scheduling algorithms do not disentangle the different aspects 
and tend to lump all of them together in the same model.

\subsubsection{Balance}
Balance requirements do not belong to the simplified model of equation 
\eqref{eqn:BasePlantModel}, which refers to a mono-processor system whose only 
resource is CPU time. It is straightforward, however, to extend the model along 
the same lines to accommodate additional resources, such as another CPU or 
I/O bandwidth. New variables would model the usage of these further resources, 
with the same assumptions as in \eqref{eqn:BasePlantModel}.
Of course, these control variables must be measurable in the real system for 
the scheduler to be effectively implementable (see \cite{HellersteinEtAl-2005a} 
for a discussion of this orthogonal aspect).
Then, control-theoretical techniques --- such as optimal control theory or 
model-predictive control --- can be used to design a scheduler which enforces 
a resource occupation given as a set point.

\subsubsection{Throughput maximization}
If throughput is part of the requirements for our scheduler, we include the 
following set of equations in the model \eqref{eqn:BasePlantModel}:
\begin{equation}
 \rho_p(k) = \max \left(\rho_p(k-1) - \tau_p(k-1),\, 0 \right)
 \label{eqn:rhoequation}
\end{equation}
Equation \eqref{eqn:rhoequation} defines $\rho_p(k)$, the remaining execution 
time of task $p$ at round $k$, as the difference between the remaining time 
during the previous round and the actual running time of $p$ during the current 
round.
Throughput maximization can then be defined as the round-wise maximization of 
the number of processes whose $\rho_p$ value is zero.
Standard control-theoretical techniques can then design a controller that 
provably achieves this requirement.

\subsubsection{Responsiveness} 
The model \eqref{eqn:BasePlantModel} includes a variable $\tau_r$ that describes
the duration of a round, hence requirements on the response time can be 
expressed as a target value for $\tau_r$. More precisely, the smaller $\tau_r$,
the more responsive is the controlled system.

\subsubsection{Other requirements}
The same framework can address other requirements, such as turnaround time, 
CPU utilization, predictability, proportionality, and deadline enforcement.
As an example, the experiments in Section \ref{sec:experimental} will address
proportionality and deadline enforcement explicitly.

\subsection{Complexity parameters}
\label{sec:complexity}
Analyzing the complexity of scheduling algorithms is often arduous, mostly 
due to the difficulty of determining the right level of abstraction to describe
the various components (i.e., the processor, the scheduler, etc.).
It also does not make sense to compare directly the general framework we
have outlined to existing algorithms; on the contrary, specific implementations
can be experimentally evaluated.

It is nonetheless interesting to present a few simple rules of thumb to have a
rough estimate of the complexity of an algorithm designed within our framework.
With the goal of determining the number of elementary operations spent by the
CPU to execute the scheduling algorithm itself, let us introduce the constants
$t_{\Sigma}$, $t_S$, $t_{\Pi}$, and $t_{\rightarrow}$.
They denote the (average) duration of a sum, subtraction, multiplication, and
bit-shift operation, respectively.
Also, let $t_c$ denote the (average) duration of a ``light'' context switch (i.e, the time overhead taken by operations such as storing and restoring context information, which does not include the actual computation of the next task to be run and its budget). Using these figures, Section \ref{sec:experimental} evaluates the 
complexity of a specific algorithm that validates our framework.

\section{Application and experimental results}
\label{sec:experimental}
This section instantiates the framework proposed by developing a 
scheduler with certain proportionality and deadline meeting requirements 
with control-theoretic techniques.
The design is evaluated on the Hartstone \cite{Hartstone-1992a,Weiderman-1989a}
benchmark, a standard real-time system benchmark.
The Hartstone benchmark evaluates deadline misses and was initially conceived 
to assess architectures and compilers, but it can be used also for evaluating 
the performances of a scheduling algorithm.

The design and the evaluation are necessarily preliminary, and do not tackle every aspect that is relevant in real-time scheduling (for example, earliness/tardiness bounds are not considered); these experiments are meant as a feasibility demonstration of the approach and tackling more challenging problems belongs to future work.

\textbf{Regulating round duration and CPU distribution.}
One of the requirements fixes a desired duration for scheduling rounds; 
let $\tau_r^{\circ}$ denote such a duration.
Moreover, define
\begin{equation}
 \theta_p^{\circ} \in \Re^N, \qquad \theta_{p,i}^{\circ} \geq 0,
 \qquad\sum\limits_{i=1}^N \theta_{p,i}^{\circ}=1
 \label{eqn:thetapo}
\end{equation}
as the vector with the \emph{fractions} of CPU time to be allotted to each task.
This vector can be expressed as a function of workload and round duration, and
the corresponding requirement be expressed as a set point for each task.
More generally, notice that requirements on fairness, tardiness, and similar 
features, are also expressible in terms of $\tau_r^{\circ}$ and 
$\theta_p^{\circ}$.
\begin{figure*}[t]
 \centering
 \includegraphics[width=0.8\textwidth]{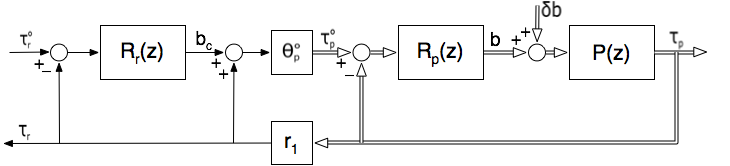}
 \caption{The control scheme proposed.}
 \label{fig:ProposedScheme-TwoLoopsLinear}
\end{figure*}

Let us now show a possible approach to design a scheduler that meets the 
requirement on the duration of the scheduling rounds.
Consider for example a cascade controller such as in Figure 
\ref{fig:ProposedScheme-TwoLoopsLinear}.
An appropriate choice for the involved regulators is to give $R_r$ a 
PI structure:
\begin{equation}
R_r(z)=k_{rr}\frac{z-z_{rr}}{z-1}
\end{equation}
while selecting $R_p$ as a diagonal integral regulator with gain $k_{pi}$:
\begin{equation}
 \begin{array}{c}
 A_{R_p}=I_N, \qquad B_{R_p}=k_{pi}I_N, \\
 C_{R_p}=I_N, \qquad D_{R_P}=0_{N\times N}.
 \end{array}
\end{equation}
Correspondingly, one can perform a model-free synthesis 
getting the values $k_{rr}=1.4$, $z_{rr}=0.88$, and $k_{pi}=0.25$. 
These values instantiate a cascade controller which we take as our BCC.
For the PSC, let us choose a simple approach where every task with a 
positive computed budget is activated following a Round Robin policy.

\textbf{Benchmark description.}
The Hartstone benchmark defines various series of tests.
Each of them starts from a 
baseline system, verifies its correct behavior, and then iteratively adds
workload and re-verifies its correct behavior until a failure occurs.
The final amount of (supported) additional workload gives a measure the system
performance. For brevity (and without loss of generality), we consider 
only the first series of the Hartstone benchmark --- the ``PH series'', 
which deals with periodic tasks --- in this paper.
The PH (Periodic tasks, Harmonic frequencies) series adds 
tasks and modifies their period and workload to stress the system.
Tasks are periodic and harmonic.

The baseline system \cite{Hartstone-1992a} consists
of five periodic tasks. Each task has a frequency and a workload.
All frequencies are an integral multiple of the smallest, and the workload is 
determined by a fixed amount of work to be completed within the task's period.
More precisely, each task has to
execute a given number of ``Wheatstones'' within a period, hence the workload 
rate is measured in Kilo-Whets instruction per second [KWIPS]. In our tests, 
we assume that the CPU can complete 1 KWIPS in 25 time units.
We do not change this figure throughout our simulations, thus neglecting the 
overhead on the hardware of executing the scheduler. In addition, let a 
frequency of 1 Hertz correspond to a period of 20000 time units.
All the tasks are independent: their execution
does not need synchronization and they are all scheduled to start at the same 
time. The deadline for the workload completion for each task coincides with the 
beginning of the next period. These assumptions are appropriate, for example, 
for programs that monitor several sensors at different rates, and display the 
results without user intervention or interrupts.
Table \ref{tab:baselineset} gives details on the baseline system.

\begin{table}[ht]
 \begin{footnotesize}
    \centering
    \begin{tabular}{|c|c|c|c|}
    \hline
    \textbf{Task} & \textbf{Frequency} & \textbf{Workload}      & \textbf{Workload rate} \\
    \hline
    \hline
    1                & 2 Hertz            & 32 Kilo-Whets          & 64 KWIPS \\
    2                & 4 Hertz            & 16 Kilo-Whets          & 64 KWIPS \\
    3                & 8 Hertz            &  8 Kilo-Whets          & 64 KWIPS \\
    4                & 16 Hertz           &  4 Kilo-Whets          & 64 KWIPS \\
    5                & 32 Hertz           &  2 Kilo-Whets          & 64 KWIPS \\
    \hline
    \end{tabular}
    \caption{The baseline task set.}
    \label{tab:baselineset}
  \end{footnotesize}
\end{table}

\textbf{Benchmark evaluation.}
We run the benchmark with three different algorithms: the one regulating CPU
distribution and round duration (designed above within our framework) with 
three different values for $\tau_r^{\circ}$, as well as the standard (real-time)
policies EDF and LLF. The yardstick for the evaluation is a simple Round Robin 
scheduler.

The presented results used the Scilab environment \cite{scilab} to perform the 
simulations; this allowed a high-level evaluation of the scheduling algorithms 
that is not tied down to any lower-level implementation detail.
As a further validation, we also run the same tests within the Cheddar framework
\cite{Cheddar-2009a}.
The results of the two sets of tests, with Cheddar and with Scilab, essentially 
coincide, therefore reinforcing our confidence in the soundness of the 
evaluation.

In the first PH test, the highest-frequency task (task 5) has the
frequency increased by 8 Hertz at each iteration, until a deadline is missed.
This tests the interactivity of the system or, in other words, its ability
to switch rapidly between tasks. In the second test, all the frequencies
are scaled by 1.1, 1.2, $\ldots$ at each iteration, until a deadline is
missed. This is a uniform increase of the number of operations done by the
tasks, therefore testing the ability to handle an
increased but still balanced workload. The third test starts from the
baseline set and increases the workload of each task by 1, 2, $\ldots$
KWPIS per period at each iteration, until a deadline is missed. This increases 
the system's overhead while introducing unbalance. In the last test, a new task 
is added at each iteration, with a workload of 8 KWPIS per period
and a frequency of 8 Hertz (equivalent to the third task of the baseline
set). This test evaluates the performance in handling a large number of
tasks.

\begin{table*}[ht]
 \begin{footnotesize}
 \begin{center}
  \begin{tabular}{|cc|cc|cc|cc|cc|}
    \hline
    & & \multicolumn{8}{c}{\textbf{Benchmark No.}} \vline \\
    & & \multicolumn{2}{c}{I}      & \multicolumn{2}{c}{II}   &
        \multicolumn{2}{c}{III}    & \multicolumn{2}{c}{IV} \vline  \\ \hline \hline
    & Period duration & \multicolumn{2}{c}{10000} \vline & \multicolumn{2}{c}{4000} \vline 
                      & \multicolumn{2}{c}{10000} \vline & \multicolumn{2}{c}{10000} \vline \\ \hline \hline
    \multirow{8}{*}{\begin{sideways}\textbf{Policy}\end{sideways}} 
    & EDF                             & 14 & (265)  & 24 & (42)   &  7 & (43)   &  7  & (73)   \\
    & LLF                             & 14 & (993)  & 24 & (1183) &  7 & (491)  &  7  & (7143) \\
    & RR, $q/T_{\min}^{base}=1/625$   &  3 & (3485) & 24 & (3999) &  3 & (4867) &  7  & (9351) \\
    & RR, $q/T_{\min}^{base}=5/625$   &  3 & (705)  & 24 & (799)  &  3 & (981)  &  7  & (1870) \\
    & RR, $q/T_{\min}^{base}=10/625$  &  3 & (357)  & 24 & (399)  &  2 & (435)  &  7  & (935)  \\
    & PSC+BCC, $\tau_r^{\circ}=500$   & 14 & (126)  & 24 & (60)   &  7 & (126)  &  7  & (252)  \\
    & PSC+BCC, $\tau_r^{\circ}=1000$  & 14 & (66)   & 24 & (36)   &  7 & (66)   &  7  & (132)  \\
    & PSC+BCC, $\tau_r^{\circ}=2000$  & 14 & (48)   & 24 & (24)   &  7 & (42)   &  7  & (84)   \\
    \hline
  \end{tabular}
 \end{center}
 \end{footnotesize}
 \caption{Hartstone PH (Periodic Tasks, Harmonic Frequencies) series benchmark: number of iterations before first deadline miss, and (in parentheses) number of context switches in the last period of the last test iteration before that with the first miss. In the RR case the quantum $q$ is selected as a fraction of the minimum task period ($625$ time units) in the baseline system, denoted by $T_{\min}^{base}$.}
 \label{tab:BenchPH-IterationsBefore1stMiss}
\end{table*}

Table \ref{tab:BenchPH-IterationsBefore1stMiss} shows the results of the PH
tests. The scheduling algorithm designed within our framework shows consistently
good performances, and can outperform other standard algorithms in certain 
operational conditions for aspects such as deadline misses.

\textbf{Complexity evaluation.}
Let $\sigma_{POL}$ denote the time spent during one round in running the
scheduler $POL$.
In our experiments, $POL$ is one of $RR$, $SRR$ (Selfish
Round Robin\footnote{Notice that the SRR is a useful example as it provides an adaptation mechanism.}), 
and $PSC+BCC$.

\begin{small}
\begin{equation}
 \begin{array}{rcl}
 \sigma_{RR}      &=& N\cdot t_{\rightarrow}+
                      N\cdot t_c, \\
 \sigma_{SRR}     &=& N\cdot t_{\rightarrow}+N\cdot t_c+
                      N^2\cdot \left(t_S+t_{\Pi}\right), \\
 \sigma_{PSC+BCC} &=& N\cdot t_{\rightarrow}+N\cdot 
                      t_c+(N+1)\cdot t_s + \\
                  & & (2N+1)\cdot t_{\Sigma}+(2N+2)\cdot t_{\Pi}
 \end{array}
\end{equation}
\end{small}

The expressions above take into account the arithmetic
operations necessary to execute the controller's code.
Then, if we denote the quantum (where applicable) by $q$, the total duration 
of one round is given by

\begin{small}
\begin{equation}
 \tau_{r,RR}      = N\cdot q, \quad
 \tau_{r,SRR}     = N_w\cdot q, \quad
 \tau_{r,PSC+BCC} = \tau_r^{\circ}
\end{equation}
\end{small}

\noindent where $N_w \leq N$ is the number of tasks in the waiting queue in the
SRR case.

Correspondingly, the overall time complexity of the algorithms can be computed.
With PSC, it is independent of the number of tasks and can be tuned by changing
the round duration parameter.
In addition, it is interesting to compare the complexity of our $PSC+BCC$ algorithm against the $RR$ algorithm (an open-loop policy) and the $SRR$ algorithm (a closed-loop variant of $RR$, where the possibility of moving tasks between queues provides the feedback mechanism).
It turns out that our $PSC+BCC$ algorithm is computationally slightly more complex than $RR$; however, the more complex properties that $PSC+BCC$ can guarantee --- such as convergence to the desired distribution in the presence of disturbances --- pay off the additional cost.
$SRR$, on the other hand, can enforce similar properties and has a greater computational complexity than $PSC+BCC$.
The comparison with $SRR$ requires, however, further investigation, because the parameters of $SRR$ do not seem to have an entirely clear interpretation within the control-theoretical framework.

\section{Conclusion and future work}
\label{sec:conclusions}
We presented a framework, based on control theory, to approach the scheduling
problem. The approach clearly separates the models of the processor and of the
scheduler. This enables the re-use of a vast repertoire of control-theoretical
techniques to design a scheduling algorithm that achieves certain requirements.
Algorithm design is then essentially reduced to controller synthesis.
We showed how to compare the resulting algorithms to existing ones, and the 
advantages that are peculiar to our approach.

This paper focused on developing the components responsible for the computation 
of budgets and the selection of tasks.
Future work will focus on the design of the intervention policies.
This aspect can still be approached within the same framework, by analyzing 
the effects of different policies on the model equations and on the overall 
system performance.
Moreover, we plan to refine the complexity evaluation of scheduling algorithms.


\end{document}